\documentclass[standard]{aa}
\usepackage{graphicx}

\begin{document}

\title{Accurate rest frequencies of methanol maser \\
and dark cloud lines\thanks{Tables 1 to 3 are also available
in electronic form at the CDS via anonymous ftp to
cdsarc.u-strasbg.fr (130.79.128.5)
or via http://cdsweb.u-strasbg.fr/cgi-bin/qcat?J/A+A/}}
\author{H. S. P. M\"uller \inst{1}, K. M. Menten \inst{2}, \and
        H. M\"ader \inst{3}}
\institute{I. Physikalisches Institut, Universit\"at zu K\"oln,
Z\"ulpicher Strasse 77, 50937 K\"oln, Germany \and
Max-Planck-Institut f\"ur Radioastronomie,
Auf dem H\"ugel 69, 53121 Bonn, Germany, \and
Institut f\"ur Physikalische Chemie, Christian-Albrechts-Universit\"at
Kiel, Olshausenstrasse 40, 24098 Kiel, Germany}
\offprints{H. S. P. M\"uller; hspm@ph1.uni-koeln.de}
\date{Received 01 June 2004 / Accepted}
\titlerunning{Methanol maser line frequencies}
\authorrunning{M\"uller et al.}

\def\meth{CH$_3$OH}
\def\kms{km\,s$^{-1}$}
\def\aap{A\&A}

\abstract{We report accurate laboratory measurements of selected
methanol transition frequencies between 0.834 and 230~GHz in order to
facilitate astronomical velocity analyses.
New data have been obtained between 10 and 27~GHz
and between 60 and 119~GHz.
Emphasis has been put on known or potential interstellar maser lines
as well as on transitions suitable for the investigation of
cold dark clouds.
Because of the narrow line widths ($<0.5$~km~s$^{-1}$) of maser lines
and lines detected in  dark molecular clouds, accurate frequencies are
needed for comparison of the velocities of different methanol lines
with each other as well as with lines from other species.
In particular, frequencies for a comprehensive set of transitions
are given which, because of their low energy levels
($< 20$~cm$^{-1}$ or 30~K), are potentially detectable in cold clouds.
Global Hamiltonian fits generally do not yet yield the required accuracy.
Additionally, we report  transition frequencies for other lines 
that may be used to test and to improve existing Hamiltonian models.}

\maketitle

\keywords{Masers -- Molecular data -- Method: laboratory --
Techniques: spectroscopic -- ISM: molecules  -- Radio lines: ISM}

\section{Introduction and Motivation}
\subsection{Methanol masers}
Methanol (\meth) is the molecule with the largest number of known
interstellar maser lines. These transitions arise from a wide range of
energy levels above the ground state, and modeling their excitation
provides important information on their emitting regions.
Based on the small number of lines known at the time,
Batrla et al. (1986) were the first to suggest that \textit{all}
methanol maser transitions belong to either one of two types,
named class I and class II by Menten (1991a,b).

Obviously, the lines considered (within each class) have to arise
from the same region for excitation modeling to make sense.
The first clue that this indeed is the case comes from the observation
that different transitions cover very similar LSR velocity ranges.
Moreover, in  particular for class I maser lines, the frequently very
sparse spectra (sometimes showing a single strong feature only) look
very similar for different transitions (apart from their intensities);
see, e.g., Fig.~1 of Menten 1991b.
The conclusion made in the latter reference,
that class I methanol masers are associated with interstellar outflows
and frequently found far away (up to a pc) from their exciting sources,
while class II masers arise from the closest, warm ($T\sim  150$~K),
dense ($n\sim 10^{6-7}$~cm$^{-3}$) environment  of newly formed
high-mass stars, has been confirmed by all subsequent observations
(see astronomical references in Table 1).
This dichotomy is coupled to their excitation: As model calculations
show (Leurini et al. 2004a), class I masers arise from basic properties
of the methanol molecule via collisional excitation, if 
no strong far-infrared field is present, i.e. far away from stars.
In contrast, class II masers are pumped by infrared radiation 
(from the embedded objects), most likely
via one or more torsionally excited states (e.g., Sobolev \& Deguchi 1994).

A common origin of different maser lines has been \textit{directly}
proven -- so far -- in only a few cases:
Very Long Baseline Interferometry observations of the  two strongest
class II maser lines, the $5_{1}\to6_{0}A^{+}$ and $2_{0}\to3_{-1}E$
transitions, at 6.7 and  12.2~GHz, respectively,
show that, indeed, maser spots in both lines do arise from identical
locations on milliarcsecond (few AU) scales  (Menten et al. 1992).
More recently, Kogan and Slysh (1998) found something similar
for class I maser transitions in DR21(OH) by combining their
observations of $7_{0}\to6_{1}A^{+}$ at 44.0~GHz with those
of $8_{0}\to7_{1}A^+$ at 95.2~GHz by Plambeck and Menten (1990).

While such direct imaging certainly proves the coexistence issue,
it is available for few lines and few sources only.
Thus, a lot of the interpretation depends on comparison of line profiles.
Therefore, a need for rest frequency values of the highest possible
accuracy is rather obvious.

\subsection{Methanol in dark clouds}
As mentioned above, methanol is readily found in hot, dense
molecular cores in high-mass star-forming regions.

In contrast, few observations of methanol in cold dark clouds 
have been made. This is partially due to the weakness of the 
emission, which is optically thin and has a low excitation temperature
in such sources, where small 
CH$_3$OH/H$_2$ ratios
of order $10^{-8}$ -- $10^{-9}$ are found. Such values can  
be readily explained 
by gas phase chemistry models. 

The first observations of CH$_3$OH in cold, dark clouds where made
by Friberg et al. (1988), who detected the $k=0$ and $1~E$-type
and the $k=0~A^+$-type lines of the $2_k - 1_k$ quartet near
96.7~GHz as well as the 48.4~GHz $1_0-0_0 A^+$ transition.
Sources included the well-studied molecule-rich clouds TMC-1,
L134N (= L183), and B335. Turner (1998) added a number of other
lines to the still small list of methanol lines observed in
dark clouds (see Table 2).

Of particular importance in this context is the detection of enhanced
absorption (over-cooling or anti-inversion) of the 12.2~GHz
$2_0-3_{-1}E$ line against the cosmic microwave background radiation
toward TMC-1 and L183 (Walmsley et al. 1988). No line better
demonstrates the class I/II maser dichotomy than this transition,
the second most luminous and widespread class II maser line known
(Batrla et al. 1987), which is pumped by prodigious far-infrared
radiation near newly formed high-mass (proto)stars.  In class I
sources, in contrast, as well as in dark clouds, i.e. in regions
devoid of strong far-infrared radiation, this line becomes
anti-inverted by the same mechanism, that produces the
$J_{k=-1} - (J-1)_{k=0}E$ ($J = 4, \ldots 9$) maser lines
(see Table 1 and Leurini et al. 2004a,b).

\section{Existing data and new laboratory measurements}

Modern millimeter-wavelength spectroscopy of methanol 
effectively started with the classic
paper by \cite{Lees1968}.

An early  compilation of line rest frequencies was presented by
\cite{Lees1973}.
Many of the millimeter and submillimeter transition frequencies commonly
used by astronomers today are the result of a concerted laboratory
measurement/theory
effort by the Duke, then Ohio State University group.
They used large quantities of measured line frequencies to fit
the (many) parameters of an extended internal axis method
Hamiltonian, yielding better than 100~kHz accuracy fits
to the bulk of the frequencies
(Herbst et al. 1984; De Lucia et al. 1989; Anderson et al. 1990).

The published frequencies in the older references had
typical uncertainties of 0.1~MHz.
At 100~GHz, this corresponds to a velocity uncertainty of 0.3~\kms,
comparable to the line widths of single maser features.

More recently, these measurements have been extended to selected 
frequency regions
between 550 and 1200~GHz by Belov et al. (1995) employing the Cologne
Terahertz Spectrometer.
Still higher frequency measurements were published e.\,g. by
Matsushima et al. (1994) and Odashima et al. (1995).
In the latter laboratory, rather comprehensive measurements
have been performed between 7 and 200~GHz (Tsunekawa et al. 1995).
Even further measurements and often model Hamiltonians
have been reported in references cited in these articles
and the ones mentioned below.

The highest accuracy measurements available for lines in the radio
frequency and centimeter wavelength ranges,
where high accuracy is most important, were made by
Radford (1972), Heuvel and Dymanus (1973), and Gaines et al. (1974),
employing beam-maser spectroscopy. Later, Mehrotra et al. (1985) and
Breckenridge \& Kukolich (1995) employed microwave
Fourier transform (MWFT) spectroscopy.
  
In this paper, we present a compilation of frequencies of all
known class I and II methanol maser lines.
Values with the highest accuracy published were chosen in general.
In addition, we used spectrometers available in Cologne and Kiel
for accurate measurements of selected methanol transition frequencies.

\section{Cologne and Kiel laboratory measurements}

\begin{figure*}
\begin{center}
\includegraphics[width=12cm,angle=-90]{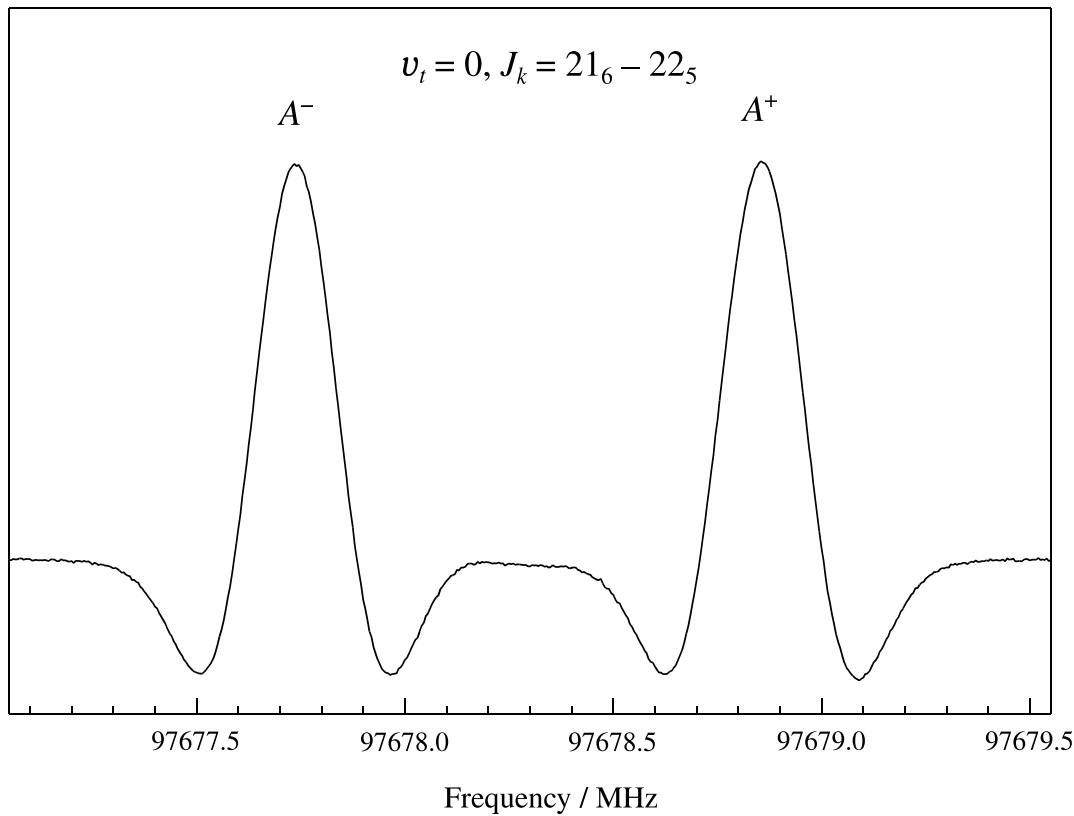}
\caption{Section of the millimeter wave spectrum of CH$_3$OH 
 showing asymmetry splitting in the $v_t = 0$, $J_k = 21_6 - 22_5$
 transitions of $A$ symmetry.
 Each line appears approximately as a second derivative of a Gaussian
 because of the $2f$-modulation employed to reduce noise and baseline 
 effects.}
\label{hot}
\end{center}
\end{figure*}

\begin{figure*}
\begin{center}
\includegraphics[width=12cm,angle=-90]{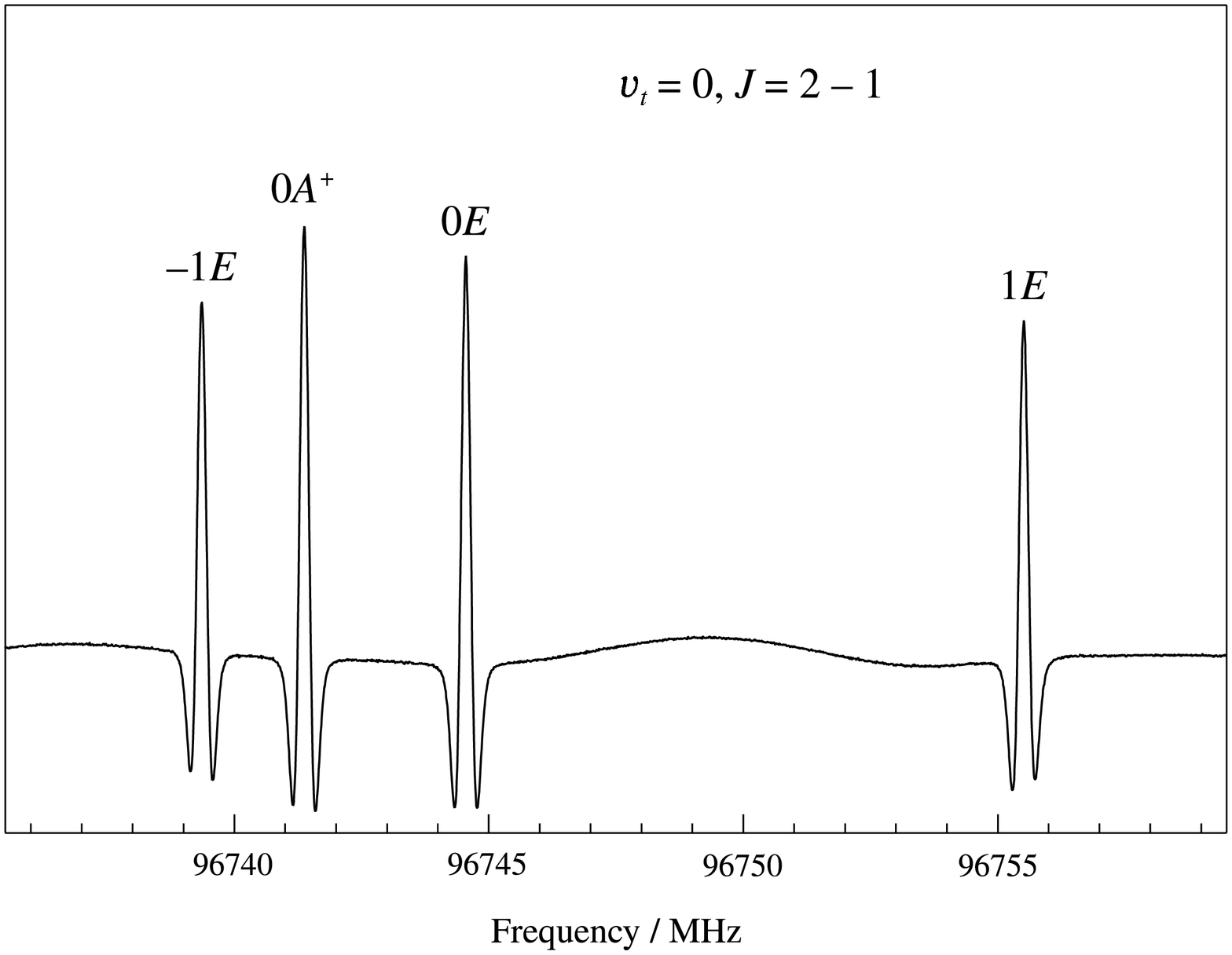}
\caption{Section of the millimeter wave spectrum of CH$_3$OH 
 showing part of the $v_t = 0$, $J = 2 - 1$ $a$-type transitions
 ($\Delta K = 0$) which are observable in dark clouds.
 The $k$ values and the symmetry are indicated.
 See also caption to Fig.~\ref{hot}.}
\label{cold}
\end{center}
\end{figure*}

The frequency region between 60 and 119~GHz has been investigated in
Cologne employing AM-MSP1 and AM-MSP2 millimeter wave
synthesizers (Analytik \&\ Me{\ss}technik GmbH,
Chemnitz, Germany) as sources and a Schottky diode detector.
Details of the spectrometer were described by Winnewisser et
al. (2000). The absorption cell was about 3.5 m long.
In order to obtain accurate line positions, low pressures
(0.1 -- 1.0 Pa) and a large number of data points were used.
Figs.~\ref{hot} and \ref{cold} demonstrate the very good
signal-to-noise ratio and the smoothness of the baseline for
strong to moderately weak transitions.

Two waveguide microwave Fourier transform spectrometers were used
for the measurements in Kiel which cover the regions of roughly
8 -- 18 and 18 -- 26~GHz. Further details on the spectrometer are given
by Kr\"uger et al. (1993) and Meyer et al. (1991).

Table~\ref{tab:mlines} presents our compilation of laboratory
frequencies of all known interstellar methanol maser lines.
References for the first astronomical detection of 
\textit{maser}\footnote{In some transitions narrow maser emission is
observed as well as broader (quasi)thermal emission.} emission
are also given for each line.
The latest version of Frank Lovas' ``NIST Recommended Rest Frequencies
for Observed Interstellar Molecular Microwave Transitions'', Revision
2002 (Lovas 2004), proved an invaluable resource in compiling it.
We augmented this Table with high accuracy rest frequencies obtained
by us and other groups where appropriate.
The methanol transition frequencies and uncertainties in the
Lovas (2004) compilation are the {\it calculated values}
from Xu \&\ Lovas (1997); see also the discussion.
The Xu \&\ Lovas (1997) data are also available online in the Cologne
Database for Molecular Spectroscopy, CDMS, described in M\"uller et al.
(2001) and M\"uller et al. (2004).

Table 2 gives frequencies for strong transitions with lower state
energies below 20~cm$^{-1}$ and frequencies below 200~GHz
for which laboratory frequencies with less than 30~kHz uncertainties
are known. Note that some transitions may be both in
Table~\ref{tab:mlines} and Table~\ref{tab:dlines}.

Only very few transitions have larger uncertainties
(of the order 50~kHz).

Finally, Table~\ref{tab:olines} summarizes the remaining transition
frequencies determined in the course of the present study. These
frequencies, together with the newly measured ones in Tables 1 and 2,
may be useful for improving existing methanol Hamiltonians.

\def\d {\phantom{$0$}}
\def\pnt {\phantom{$\nu_2=1,$}}
\def\sr { }
\def\SC {{\bf SC}}
\def\JK         {$J_{K_{\rm a},K_{\rm c}}=$}
\def\pJK        {\phantom{$J_{K_{\rm a},K_{\rm c}}=$}}
\def\Jk         {$J_k=$}
\def\pJk        {\phantom{$J_k=$}}
\def\pJ         {\phantom{$J=~$}}
\def\JKN        {$(J,K)=$}
\def\pJKN       {\phantom{$(J,K)=$}}

 \begin{center}
\begin{table*}[tb]
 \caption{Methanol maser transitions -- maser class, frequencies,
lower state energies, line strengths, and laboratory and
astronomical references}
 \label{tab:mlines}
 \begin{tabular}{rllllll}
 \hline \hline   
\noalign{\smallskip}                               
Class$^a$& Transition, $J_k=^b$& Frequency (MHz)$^c$ &  $E^d_\ell$ (cm$^{-1}$) & $S\mu^2$ (D$^2$)
& Laboratory Ref.$^e$ & Astronomical Ref.$^f$\\
\noalign{\smallskip}                               
\hline 
\noalign{\smallskip}
I?$^g$ &\d$1_1\to1_1A^{\mp}$ &\d\d\d834.267(2)$^h$ & \d11.705 & \d1.2129 & Radford 1972          & Ball et al. 1970 \\
I?$^g$ &\d$3_{1}\to3_{1}A^{\mp}$ & \d\d5005.32079(20)$^h$ & \d19.703 &\d0.4693 & HD73 & Robinson et al. 1973\\
II&\d$5_{1}\to6_{0}A^{+}$   &\d\d6668.5192(8)  & \d33.876 & \d5.0768 & BR95                & Menten 1991\\
 I&\d$9_{-1}\to8_{-2}E$     &\d\d9936.202(4)   & \d76.102 & \d2.7769 & BR95                & Slysh et al. 1993\\
II&\d$2_{0}\to3_{-1}E$      &\d12178.597(4)    & \d13.556 & \d1.9848 & BR95                & Batrla et al. 1987\\
II&\d$2_{1}\to3_{0}E$       &\d19967.3961(2)   & \d18.803 & \d0.9244 & M85                 & Wilson et al. 1985\\
II&\d$9_{2}\to10_{1}A^{+}$  &\d23121.0242(5)   & \d98.053 & \d3.1467 & M85                 & Wilson et al. 1984\\
 I&\d$3_{2}\to3_{1 }E$      &\d24928.707(7)    & \d24.310 & \d2.8073 & This work           & Barrett et al. 1975\\
 I&\d$4_{2}\to4_{1}E$       &\d24933.468(2)    & \d30.764 & \d3.9283 & Gaines et al. 1974  & Barrett et al. 1971  \\
 I&\d$2_{2}\to2_{1}E$       &\d24934.382(5)    & \d19.469 & \d1.5948 & Gaines et al. 1974  & Barrett et al. 1975\\
 I&\d$5_{2}\to5_{1}E$       &\d24959.0789(4)   & \d38.833 & \d5.0264 & M85                 & Barrett et al. 1971\\
 I&\d$6_{2}\to6_{1}E$       &\d25018.1225(4)   & \d48.514 & \d6.1287 & M85                 & Barrett et al. 1971\\
 I&\d$7_{2}\to7_{1}E$       &\d25124.8719(4)   & \d59.809 & \d7.2483 & M85                 & Barrett et al. 1971\\
 I&\d$8_{2}\to8_{1}E$       &\d25294.4165(2)   & \d72.717 & \d8.3910 & M85                 & Barrett et al. 1971\\
 I&\d$9_{2}\to9_{1}E$       &\d25541.3979(4)   & \d87.239 & \d9.5570 & M85                 & Menten et al. 1986 \\
 I&$10_{2}\to10_{1}E$       &\d25878.2661(4)   &  103.373 &  10.7398 & M85                 & Matsakis et al. 1980\\
 I&$12_{2}\to12_{1}E$       &\d26847.233(50)   &  140.478 &  13.0919 & T95                 & Wilson et al. 1996 \\
 I&$13_{2}\to13_{1}E$       &\d27472.531(30)   &  161.449 &  14.2057 & T95                 & Wilson et al. 1996 \\
 I&$14_{2}\to14_{1}E$       &\d28169.462(30)   &  184.032 &  15.2244 & T95                 & Wilson et al. 1996 \\
 I&$15_{2}\to15_{1}E$       &\d28905.812(30)   &  208.226 &  16.0977 & T95                 & Wilson et al. 1996 \\
II&\d$8_{2}\to9_{1}A^{-}$   &\d28969.942(50)   & \d83.319 & \d3.0236 & T95                 & Wilson et al. 1993 \\
 I&$16_{2}\to16_{1}E$       &\d29636.936(10)   &  234.031 &  16.7716 & T95                 & Wilson et al. 1996 \\
 I&$17_{2}\to17_{1}E$       &\d30308.034(10)   &  261.445 &  17.1961 & T95                 & Wilson et al. 1996 \\
 I&\d$4_{-1}\to3_{0}E$      &\d36169.265(30)   & \d18.803 & \d2.5184 & T95                 & Morimoto et al. 1985$^\star$ \\
II&\d$7_{-2}\to8_{-1}E$     &\d37703.700(30)   & \d61.930 & \d2.4051 & T95                 & Haschick et al. 1989 \\
II&\d$6_{2}\to5_{3}A^{+}$   &\d38293.268(50)   & \d58.813 & \d0.9488 & T95                 & Haschick et al. 1989\\
II&\d$6_{2}\to5_{3}A^{-}$   &\d38452.677(50)   & \d58.813 & \d0.9495 & T95                 & Haschick et al. 1989\\
 I&\d$7_{0}\to6_{1}A^{+}$   &\d44069.410(10)   & \d43.694 & \d6.1380 & T95                 & Morimoto et al. 1985 \\
 I&\d$5_{-1}\to4_{0}E$      &\d84521.169(10)   & \d25.254 & \d3.0830 & This work           & Batrla \& Menten 1988$^\star$ \\
II&\d$7_{2}\to6_{3}A^{-}$   &\d86615.600(5)    & \d68.493 & \d1.3578 & This work           & Sutton et al. 2001$^\star$\\
II&\d$7_{2}\to6_{3}A^{+}$   &\d86902.949(5)    & \d68.493 & \d1.3596 & This work           & Sutton et al. 2001$^\star$\\
 I&\d$8_{0}\to7_{1}A^{+}$   &\d95169.463(10)   & \d54.888 & \d7.2211 & This work           & Plambeck \&\ Wright 1988$^\star$\\
 I&$11_{-1}\to10_{-2}E$     & 104300.414(7)    &  106.779 & \d3.4141 & This work          & Voronkov et al. 2004$^\star$\\
II&\d$3_{1}\to4_{0}A^{+}$   & 107013.803(5)    & \d16.134 & \d3.0088 & This work           & Val'tts et al. 1995a$^\star$\\
II&\d$0_{0}\to1_{-1}E$      & 108893.963(7)    &\d\d5.490 & \d0.9784 & This work           & Val'tts et al. 1999$^\star$\\
 I&\d$6_{-1}\to5_{0}E$      & 132890.692(10)   & \d33.316 & \d3.6871 & T95                 & Slysh et al. 1997$^\star$\\
I&\d$9_{0}\to8_{1}A^{+}$    & 146618.794(50)   & \d67.679 & \d8.3288 & T95                 & Menten 1991b     \\
II&\d$8_{0}\to8_{-1}E$      & 156488.868(10)   & \d61.930 & \d6.7830 & T95                 & Slysh et al. 1995a$^\star$\\
II&\d$2_{1}\to3_{0}A^{+}$   & 156602.413(10)   &\d\d9.681 & \d1.9963 & T95                 & Slysh et al. 1995a$^\star$\\
II&\d$7_{0}\to7_{-1}E$      & 156828.533(10)   & \d49.035 & \d6.2749 & T95                 & Slysh et al. 1995a$^\star$ \\
II&\d$6_{0}\to6_{-1}E$      & 157048.625(10)   & \d37.749 & \d5.6636 & T95                 & Slysh et al. 1995a$^\star$\\
II&\d$5_{0}\to5_{-1}E$      & 157179.017(10)   & \d28.073 & \d4.9589 & T95                 & Slysh et al. 1995a$^\star$\\
II&\d$4_{0}\to4_{-1}E$      & 157246.056(10)   & \d20.009 & \d4.1725 & T95                 & Slysh et al. 1995a$^\star$\\
II&\d$1_{0}\to1_{-1}E$      & 157270.851(10)   &\d\d5.490 & \d1.4611 & T95                 & Slysh et al. 1995a$^\star$\\
II&\d$3_{0}\to3_{-1}E$      & 157272.369(10)   & \d13.556 & \d3.3178 & T95                 & Slysh et al. 1995a$^\star$\\
II&\d$2_{0}\to2_{-1}E$      & 157276.058(10)   &\d\d8.717 & \d2.4090 & T95                 & Slysh et al. 1995a$^\star$\\
 I&\d$8_{-1}\to7_{0}E$      & 229758.760(50)   & \d54.266 & \d5.0473 & Sastry et al. 1985  & Slysh et al. 2002$^\star$\\
\noalign{\smallskip}
 \hline
 \end{tabular}
\medskip

$^a$Gives the maser type. Interstellar methanol masers are either
class I or to II; see text and Menten 1991a,b.
$^b$The parity of $A$ symmetry transitions is given as superscript.
If the parities of the upper (left) and lower (right) state differ,
the parities are given as upper and lower superscript, respectively.
$^c$Numbers in parentheses denote the measurement uncertainties
in units of the least significant figures.
$^d$Lower state level energies are {\it all} relative to the
ground-state ($0^+_0$) level of the $A$-symmetry species.  The E-type
ground state ($1_{-1}$) level is at a 5.490~cm$^{-1}$ higher energy.
$^e$HD73 stands for Heuvel \&\ Dymanus 1973; BR95 for Breckenridge \&\
Kukolich 1995, M85 for Mehrotra et al. 1985, and T95 for Tsunekawa et
al. 1995.
$^f$Only the first reference reporting \textit{maser} emission
in this line is given.  An asterisk means that this line was earlier
detected in thermal emission or absorption.
See Lovas (2004) for relevant references.
$^g$See section 4.3. for a discussion of these lines.
$^h$The $^1$H hyperfine splitting has been resolved to some extent in
the laboratory. For the $J = 1 \to 1$ transition, four components, spread
over 19.4~km\,s$^{-1}$, have been observed. The frequency in the Table
is the one of the strongest component; the intensity weighted average
is 1~kHz lower.
For the $J = 3 \to 3$ transition, two components, split by
0.19~km\,s$^{-1}$, have been observed.
\end{table*}

\end{center}


 \begin{center}
\begin{table*}[tb]
 \caption{Accurately known methanol transitions detected or potentially
   detectable in dark clouds, frequencies, lower state energies,
   line strengths, and laboratory and dark cloud references}
 \label{tab:dlines}
 \begin{tabular}{cllllll}
 \hline \hline   
\noalign{\smallskip}                               
Maser$^{a}$&Transition, $J_k=^b$& Frequency (MHz)$^c$ & $E_\ell^d$ (cm$^{-1}$) & $S\mu^2$ (D$^2$)
& Laboratory Ref.$^e$ & Dark cloud ref.$^f$\\
\noalign{\smallskip}                               
\hline 
\noalign{\smallskip}
I?&$1_{1}\to1_{1}A^{\mp}$&\d\d\d834.267(2)    &  11.705 & 1.2129 & Radford 1972 &\textit{Slysh et al. 1995b}\\
 &$2_{1}\to2_{1}A^{\mp}$ & \d\d2502.7785(10)  &  14.904 & 0.6725 & HD73         & ND \\
I?&$3_{1}\to3_{1}A^{\mp}$& \d\d5005.32079(20) &  19.703 & 0.4693 & HD73         &\textit{Kalenski{\u i} et al. 2004}\\
II&$2_{0}\to3_{-1}E^f$   &  \d12178.595(3)    &  13.556 & 1.9848 & Gaines et al. 1974 & Walmsley et al. 1988 \\
II&$2_{1}\to3_{0}E$      &  \d19967.3961(2)   &  18.803 & 0.9224 & M85 \\
I&$2_{2}\to2_{1}E$       &  \d24934.382(5)    &  19.469 & 1.5948 & Gaines et al. 1974 \\
I&$4_{-1}\to3_{0}E$      &  \d36169.265(30)   &  18.803 & 2.5184 & T95 \\
&$1_{0}\to0_{0}A^+$      &  \d48372.4558(7)   & \d0.000 & 0.8086 & HD73        & Friberg et al. 1988 \\
&$1_{0}\to0_{0}E$        &  \d48376.892(10)   & \d9.122 & 0.8084 & T95         & \textit{Kaifu et al. 2004} \\
&$1_{0}\to2_{-1}E$       &  \d60531.489(10)   & \d8.717 & 1.4742 & This work   & NO \\
&$1_{1}\to2_{0}E$        &  \d68305.640(20)   &  13.963 & 0.4573 & This work \\
&$2_{1}\to1_{1}A^+$      &  \d95914.309(5)    &  11.705 & 1.2141 & This work \\
&$2_{-1}\to1_{-1}E$      &  \d96739.362(5)    & \d5.490 & 1.2133 & This work  & Friberg et al. 1988\\
&$2_{0}\to1_{0}A^+$      &  \d96741.375(5)    & \d1.614 & 1.6171 & This work  & Friberg et al. 1988 \\
&$2_{0}\to1_{0}E$        &  \d96744.550(5)    &  10.736 & 1.6167 & This work  & Friberg et al. 1988 \\
&$2_{1}\to1_{1}E$        &  \d96755.511(5)    &  16.241 & 1.2443 & This work  & Menten et al. 1988 \\
&$2_{1}\to1_{1}A^-$      &  \d97582.804(7)    &  11.733 & 1.2141 & This work \\
II&$3_{1}\to4_{0}A^+$    &   107013.803(5)    &  16.134 & 3.0088 & This work \\
II&$0_{0}\to1_{-1}E$     &   108893.963(7)    & \d5.490 & 0.9784 & This work   & Turner 1998 \\
&$2_{2}\to1_{1}E$        &   121689.975(10)   &  16.241 & 2.8297 & T95         \\
&$3_{1}\to2_{1}A^+$      &   143865.801(10)   &  14.904 & 2.1584 & T95         \\
&$3_{0}\to2_{0}E$        &   145093.707(10)   &  13.963 & 2.4249 & T95         \\
&$3_{-1}\to2_{-1}E$      &   145097.370(10)   & \d8.717 & 2.1569 & T95         & Turner 1998 \\
&$3_{0}\to2_{0}A^+$      &   145103.152(10)   & \d4.840 & 2.4257 & T95         & Turner 1998 \\
&$3_{1}\to2_{1}E$        &   145131.855(10)   &  19.469 & 2.2119 & T95         \\
&$3_{1}\to2_{1}A^-$      &   146368.342(50)   &  14.988 & 2.1585 & T95         \\
II&$2_{1}\to3_{0}A^+$    &   156602.413(10)   & \d9.681 & 1.9963 & T95         \\
II&$1_{0}\to1_{-1}E$     &   157270.851(10)   & \d5.490 & 1.4611 & T95         \\
II&$3_{0}\to3_{-1}E$     &   157272.369(10)   &  13.556 & 3.3178 & T95         \\
II&$2_{0}\to2_{-1}E$     &   157276.058(10)   & \d8.717 & 2.4090 & T95         \\
&$1_{1}\to1_{0}E$        &   165050.195(10)   &  10.736 & 1.3466 & T95         \\
&$2_{1}\to2_{0}E$        &   165061.156(10)   &  13.963 & 2.2373 & T95         \\
&$3_{1}\to3_{0}E$        &   165099.271(10)   &  18.803 & 3.1170 & T95         \\
&$3_{2}\to2_{1}E$        &   170060.581(10)   &  19.469 & 3.1205 & T95         & ND \\
&$4_{1}\to3_{1}A^+$      &   191810.509(10)   &  19.703 & 3.0352 & T95         & ND \\
&$4_{0}\to3_{0}E$        &   193415.367(10)   &  18.803 & 3.2327 & T95         & ND \\
&$4_{-1}\to3_{-1}E$      &   193441.610(10)   &  13.556 & 3.0330 & T95         & ND \\
&$4_{0}\to3_{0}A^+$      &   193454.361(10)   & \d9.681 & 3.2342 & T95         & ND \\
&$4_{1}\to3_{1}A^-$      &   195146.760(10)   &  19.870 & 3.0353 & T95         & ND \\
\noalign{\smallskip} \hline
\end{tabular}
\medskip

For nomenclature see also notes to Table 1.
$^a$I and II mean that these lines are class I and II maser
transitions, respectively, in some non-dark cloud regions (see Table 1).
$^f$Only references reporting the first detection of a line in cold
($T\approx 10$~K) dark clouds are listed.
For non-dark cloud astronomical references to first detections,
see Lovas (2003).  ``ND'' means that no astronomical
detection is reported in the latter reference; ``NO'' means that
the line is not observable from the ground.
$^g$This line appears in enhanced absorption (i.e. is over-cooled)
against the cosmological microwave background radiation.
References in italics report observations of a given
line that did not yield its detection.
\end{table*}

\end{center}


 \begin{center}
\begin{table*}[tb]
 \caption{Other methanol transitions measured in the course of the
present investigation}
 \label{tab:olines}
 \begin{tabular}{lllllllll}
 \hline \hline   
\noalign{\smallskip}                               
$v_t$& Transition, $J_k=^b$& Frequency (MHz)$^c$ & & & $v_t$& Transition, $J_k=^b$& Frequency (MHz)$^c$ & \\
\noalign{\smallskip}                               
\hline 
\noalign{\smallskip}
     0 & \d$4_{3}\to5_{2}A^-$      & \d10058.268(2)       &    & &   1 & \d$2_{-1}\to1_{-1}E$      & \d96501.705(10)      &    \\
     0 & $32_{2}\to32_{1}E$        & \d11841.902(1)       & ND & &   1 & \d$2_{0}\to1_{0}A^+$      & \d96513.675(10)      &    \\
     0 & $31_{2}\to31_{1}E$        & \d14407.777(1)       & ND & &   1 & \d$2_{1}\to1_{1}A^-$      & \d96588.582(5)       &    \\
     0 & $32_{8}\to31_{9}E$        & \d14446.665(1)       & ND & &   0 & $21_{6}\to22_{5}A^-$      & \d97677.738(5)       &    \\
     0 & $30_{2}\to30_{1}E$        & \d16941.190(1)       & ND & &   0 & $21_{6}\to22_{5}A^+$      & \d97678.857(7) &    \\
     0 & \d$6_{1}\to6_{1}A^{\mp}$  & \d17513.342(2)x      & ND & &   0 & $15_{1}\to15_{1}A^{\mp}$  & \d99602.078(10)      & ND \\
     0 & $29_{2}\to29_{1}E$        & \d19390.121(6)       & ND & &   0 & \d$9_{-2}\to9_{1}E$       &  101737.173(20)      &    \\
     1 & $10_{1}\to11_{2}A^+$      & \d20970.651(1)       &    & &   0 & $10_{-2}\to10_{1}E$       &  102122.667(10)      &    \\
     1 & $12_{2}\to11_{1}A^-$      & \d21550.300(6)       &    & &   0 & $11_{-2}\to11_{1}E$       &  102658.046(7)       &    \\
     0 & $28_{2}\to28_{1}E$        & \d21708.662(5)       & ND & &   0 & $22_{-2}\to22_{2}E$       &  102704.298(20)      & ND \\
     0 & $27_{2}\to27_{1}E$        & \d23854.261(6)       & ND & &   1 & $15_{-2}\to16_{-3}E$      &  102957.691(7)       &    \\
     0 & $26_{2}\to26_{1}E$        & \d25787.125(10)      & ND & &   0 & $12_{-2}\to12_{1}E$       &  103381.146(5)       &    \\
     1 & $10_{1}\to11_{2}A^-$      & \d26120.584(10)      & ND & &   0 & $13_{-3}\to12_{-4}E$      &  104060.657(10)      &    \\
     0 & \d$8_{2}\to9_{1}A^+$      & \d66947.895(20)      & ND & &   0 & $13_{-2}\to13_{1}E$       &  104336.562(7)       &    \\
     0 & $11_{1}\to10_{2}A^-$      & \d76247.283(5)       &    & &   0 & $10_{4}\to11_{3}A^-$      &  104354.819(5)       &    \\
     0 & \d$5_{0}\to4_{1}E$        & \d76509.668(7)       &    & &   0 & $10_{4}\to11_{3}A^+$      &  104410.446(5)       &    \\
     0 & $15_{-2}\to15_{2}E$       & \d78254.011(20)      & ND & &   0 & $13_{1}\to12_{2}A^+$      &  105063.691(5)       &    \\
     0 & \d$7_{2}\to8_{1}A^-$      & \d80993.241(7)       &    & &   0 & $14_{-2}\to14_{1}E$       &  105576.290(10)      &    \\
     0 & $17_{-2}\to17_{2}E$       & \d81318.424(20)      & ND & &   0 & $15_{-2}\to15_{1}E$       &  107159.820(10)      &    \\
     0 & $18_{-2}\to18_{2}E$       & \d83792.561(10)      & ND & &   0 & $26_{0}\to26_{-1}E$       &  109136.883(7)       &    \\
     0 & $13_{-3}\to14_{-2}E$      & \d84423.776(5)       &    & &   0 & $16_{-2}\to16_{1}E$       &  109153.107(10)      &    \\
     1 & $13_{10}\to13_{11}E$      & \d84930.946(10)      & ND & &   0 & $23_{-2}\to23_{2}E$       &  110060.492(20)      & ND \\
     1 & $14_{10}\to14_{11}E$      & \d85346.076(7)       & ND & &   0 & $18_{0}\to17_{3}E$        &  111254.256(20)      & U  \\
     0 & \d$6_{-2}\to7_{-1}E$      & \d85568.084(10)      &    & &   0 & \d$7_{2}\to8_{1}A^+$      &  111289.550(10)      &    \\
     1 & $15_{10}\to15_{11}E$      & \d85773.092(7)       & ND & &   0 & $17_{-2}\to17_{1}E$       &  111626.449(15)T     &    \\
     0 & $19_{-2}\to19_{2}E$       & \d87066.748(10)      & ND & &   0 & $14_{-3}\to15_{2}E$       &  112490.953(15)      & ND \\
     0 & $14_{1}\to14_{1}A^{\mp}$  & \d87241.339(10)      & ND & &   0 & $16_{1}\to16_{1}A^{\mp}$  &  112745.243(10)      & ND \\
     0 & \d$8_{-4}\to9_{-3}E$      & \d89505.808(5)       &    & &   0 & $25_{0}\to25_{-1}E$       &  113408.094(7)       & ND \\
     0 & $20_{-2}\to20_{2}E$       & \d91254.686(10)      & ND & &   0 & $18_{-2}\to18_{1}E$       &  114650.861(5)       &    \\
     1 & \d$1_{0}\to2_{1}E$        & \d93196.673(10)      &    & &   0 & $19_{-3}\to20_{0}E$       &  115800.166(20)      & ND \\
     1 & $10_{-7}\to11_{-6}E$      & \d94022.734(10)      & ND & &   0 & $24_{0}\to24_{-1}E$       &  117636.228(10)      & NO \\
     0 & \d$8_{3}\to9_{2}E$        & \d94541.765(5)       & ND & &   1 & $10_{2}\to9_{3}E$         &  118156.280(10)      & NO \\
     0 & $19_{7}\to20_{6}A$        & \d94814.987(10)      & ND & &   1 & \d$8_{1}\to9_{2}A^+$      &  118207.927(5)       & NO \\
     1 & \d$2_{1}\to1_{1}A^+$      & \d96396.040(7)       &    & &   0 & $19_{-2}\to19_{1}E$       &  118293.492(5)       & NO \\
     0 & $21_{-2}\to21_{2}E$       & \d96446.619(15)      &    & &   1 & $17_{6}\to18_{5}A$        &  118481.011(5)       & NO \\
     1 & \d$2_{1}\to1_{1}E$        & \d96492.152(7)       &    & &   0 & $24_{-2}\to24_{2}E$       &  118522.453(15)      & NO \\
     1 & \d$2_{0}\to1_{0}E$        & \d96493.540(10)      &    & &     &                           &                      &     \\
\noalign{\smallskip}
 \hline
 \end{tabular}\medskip

For nomenclature see notes to Table 1.
Tsunekawa et al. (1995) give 111626.846(30) MHz for the frequency
of the line marked with T; we suspect a typographical error,
see section 4.1, where also some assignments for $J > 30$ are discussed.
Heuvel and Dymanus 1973 give 17513.34127(20)~MHz for the unsplit
position of the line  marked with x.
All lines, except those marked ``ND''  have been
detected in astronomical sources.
See Lovas (2004) for references to first detections.
``ND'' means that no astronomical detection is reported
in the latter reference.
``U'' means that this line is listed as unidentified in Lovas (2004).
\end{table*}


\end{center}

\section{Discussion and conclusions}

\subsection{Comparison with earlier results}

The newly obtained transition frequencies are generally much
more accurate than previously published values, sometimes by
more than a factor of 10. Moreover, they agree generally
well within experimental uncertainties both with older
measurements summarized in Xu \&\ Lovas (1995) {\it and}
with more recent measurements from Tsunekawa et al. (1995).
Specifically, among the new lines in Tables 1 and 2 there is only
one for which the agreement is not quite within estimated
uncertainties with either data. The value  of 86615.760~MHz in
Xu \&\ Lovas (1995) originates from Sastry et al. (1985) and
has an uncertainty of 100 kHz in that paper.
Tsunekawa et al. (1995) give 86615.578~MHz with an uncertainty
of 10~kHz. Both values differ slightly from our value of
86615.600(5)~kHz. But since experimental uncertainties are
usually given as $1\sigma$ values, agreement ``only''
within twice or three times the larger uncertainties
should not be of major concern.
Moreover, our current value is in excellent agreement with the
calculated value of 86615.602(14)~kHz in Xu \&\ Lovas (1995).
Also, the $J = 2 \to 1$ transition frequencies around 96~GHz given
in Table~\ref{tab:dlines} agree within 1 to 4~kHz with {\it calculated}
values in Xu \&\ Lovas (1995). In most instances, this is less
than {\it both} the experimental and the calculated uncertainties.

The agreement of the transition frequencies in Table~\ref{tab:olines}
with calculated values in Xu \&\ Lovas (1995) is mostly within
50~kHz or three times the calculated uncertainties,
whichever is larger;
only some high $J$ lines show larger deviations.
This level of agreement is satisfactory.
It should be pointed out that the calculated frequencies and
uncertainties not only in Xu \&\ Lovas (1995), but for all
Hamiltonian models depend on the quality of the model employed
and on the quality of the input data.
Since the input data in Xu \&\ Lovas (1995) has an accuracy of
50 or 100~kHz for the most part it cannot be expected
that the predictions are much better than this
{\it in general} -- irrespective of the predicted uncertainties.
Moreover, while predictions are {\it usually} reliable in cases
of interpolation, extrapolations should always be
viewed with caution.

In addition, most transitions in Table 3 had been measured already
by Tsunekawa et al. 1995. The agreement is almost always good
to very good, meaning within three and one times the combined
experimental uncertainties, respectively.
There are, however, lines with larger deviations:
For the $17_{-2}\to17_{1}E$ line they report 111626.846(30)~MHz
while we measured 111626.449(15)~MHz.
The latter value is in good agreement with 111626.530(100)~MHz
measured by Sastry et al. (1985) and with 111626.550(35)~MHz
calculated by Xu \&\ Lovas (1995). Therefore, we suspect a
typographical error by Tsunekawa et al. A bigger discrapancy occured
for the $32_{2}\to32_{1}E$ transition. Tsunekawa et al. (1995) report
12198.200(50)~MHz. No line was measured at that position in the
course of the present investigation. Pearson communicated an
alternative position almost 200~MHz lower for this transition
from an ongoing investigation into the ro-torsionial spectrum of
CH$_3$OH, a reassignment for the line at 14446.665~MHz,
supposedly the $31_{2}\to31_{1}E$ transition, as well as
an alternative position for the latter. Lines with very good
signal-to-noise ratio could be measured within 100~kHz
of the predictions.

\subsection{Astronomical impact}

On occasion, astronomical measurements have been used to propose
improved rest frequencies compared to the ones obtained in the
laboratory. While it may be sometimes easy to obtain very precise
astronomical line frequencies, the determination of
{\it accurate} positions (i.\,e. the absolute line position)
is significantly more difficult.
Nevertheless, astronomical observations have pointed at deficiencies
in the accuracy of laboratory rest frequencies and have provided
improved values in several instances.
In fact, the laboratory measurements by Gaines et al. (1974) were
sparked by seemingly different velocity shifts for the
$E$-symmetry, $k = 2 \to 1$, $\Delta J = 0$ transitions with
$J =$ 6 and 7 observed by Chui et al. (1974).
The greatly improved transition frequencies of Gaines et al. (1974)
resolved these discrepancies.

Friberg et al. (1988) proposed an improved rest frequency of the
$2_0-1_0 A^+$ line by comparison with the velocities of other lines
with  more accurately known frequencies. However, their shift
from 96741.42~MHz to 96741.39~MHz was close to the extrapolated
uncertainty of 0.04 MHz. On the other hand, Turner (1998) found
no need for a revised rest frequency; his downward shift by 8~kHz
was well within the uncertainty of 32~kHz.
Our new frequency implies an even larger downward shift of
45 kHz, see Table~\ref{tab:dlines}, which is in perfect agreement
with the calculated position from Xu \&\ Lovas (1997), see above.

To give just one example of the impact our (in many cases revised)
frequencies have on the interpretation of maser data, we mention that
Val'tts et al. (1995) proposed a revision of the 107013.85(10)~MHz
rest frequency by Lees \&\ Baker (1968) to 107013.67~MHz.
The present results clearly show that that shift is too large.
In fact, our value of 107013.803(5)~MHz, see Table~\ref{tab:mlines},
is well within the uncertainties of Lees \&\ Baker (1968).
Indeed, when critically examining their revised velocities, e.g. ,
that of the strongest maser feature in W3(OH) and the thermal emission
components in Orion-KL, with literature values of other methanol lines,
we believe that the value presented here is the most appropriate.

\subsection{A note on the low frequency $\Delta J =0$ lines}

The identification of certain lower-$J$, $\Delta J = 0$,
transitions as masers is uncertain.
Given the large linewidths and extended emission distribution
observed in the Galactic center region (and only there),
the $1_1A^-\to1_1A^+$ line, the first interstellar methanol
transition discovered, does clearly not show high gain maser action.
However, the fact that it appears in emission against a
high brightness temperature continuum background indicates
that it must be weakly inverted, like low quantum number transitions
from various other complex molecules found near the Galactic center;
see Menten (2004).
Optical depth estimates are of order $10^{-3}$ (see Gottlieb
et al. 1979), which implies certainly no strong maser action.
The same arguments apply to the $3_1A^-\to3_1A^+$ line,
which is also seen in emission toward Sgr B2 and at least one other
location in the Galactic center region, G$0.5-0.1$ (Sgr B1;
Mezger \&\ Smith 1976).  Recent calculations on CH$_3$OH excitation
predict these lines and other $J_kA^-\to J_kA^+$ lines to be very 
weak class I masers under certain conditions (Leurini et al. 2004a).

\section{Summary}

The present manuscript provides accurate transition
frequencies for methanol maser as well as dark cloud lines,
several of which have been determined in the course of the
present investigation. This compilation of methanol lines should be
useful for velocity analyses of methanol masers and of dark clouds.
Moreover, the newly determined transition frequencies will be useful
for improving existing Hamiltonian models for methanol.
This is of particular importance for obtaining improved methanol
rest frequencies in the 26 -- 54 GHz region for which it may be
difficult to improve the data in the laboratory.

\acknowledgements{The work in Bonn and Cologne has been supported by the
Deutsche Forschungsgemeinschaft via grant SFB 494. Additional funding by
the science ministry of the Land Nordrhein-Westfalen is also
acknowledged. We thank John C. Pearson for comments on and
new predictions for some transitions with $J > 30$.
We also thank the referee, Slava Slysh, for a super-diligent
reading and his useful remarks; in particular he pointed out a 
newly detected maser line (at 104.3~GHz).}

\end{document}